\documentclass[11pt]{article}
\usepackage[dvipdfmx]{graphicx}
\usepackage{color}
\usepackage{amsmath,amssymb,amsfonts,latexsym}
\begin{document}

\begin{center}
\textbf{\Large Emperical Study on the Effect of Multi-Sampling\\[2mm] in the Prediction Step of the Particle Filter}

\vspace{8mm}
{\large Genshiro Kitagawa}\\[2mm]
The Institute of Statistical Mathematics\\[-1mm]
and\\[-1mm]
Graduate University for Advanced Study

\vspace{3mm}
{\today}
\end{center}

% \maketitle

\noindent
\textbf{Abstract:}

Particle filters are applicable to a wide range of nonlinear, non-Gaussian state-space models and have already been applied to a variety of problems.
However, there is a problem in the calculation of smoothed distributions, where particles gradually degenerate and accuracy is reduced.
The purpose of this paper is to consider the possibility of generating multiple particles in the prediction step of the particle filter and to empirically verify the effect using real data.

\vspace{2mm}

\section{Introduction}
In time series analysis, the prior knowledge of the dynamics of the phenomena and the mechanism of the observation process can usually be combined into state-space model form.
And many important problems in time series analysis can be solved using state-space models (Harrison and Stevens 1976, West and Harison 1989, Kitagawa and Gersch 1996, Doucet et al. 2001, Prado and West 2010, Kitagawa and Gersch 1984).
In the 1990's, various sequential Monte Carlo methods, referred to as bootstrap filters, Monte Carlo filters, and particle filters, were developed (Gordon et al.~1993, Kitagawa 1993, 1996, Doucet et al.~2000, 2001).
In these methods, arbitrary distributions of the state and the system noise are expressed by many particles. 
Then, it is possible to develop a recursive filter and smoother for general nonlinear non-Gaussian state-space models.
These methods have been successfully applied to a number of complex real-world problems (Doucet et al. 2001).

In state-space modeling, it is important to compute smoothed distributions that make good use of all the data already obtained.
However, the usual method used in the calculation of smoothed distributions preserves particles that represent past filter distributions and resamples them using weight coefficients obtained from new observations, but if this is repeated, the weights are concentrated on specific particles, the distribution becomes degenerate, and the accuracy of the distribution approximation may deteriorate rapidly. The accuracy of the approximation of the distribution can deteriorate eventually.
Therefore, it is important to develop a method that is less likely to cause degeneracy of the distribution. In this paper, we perform an emperical study using a specific trend estimation as an example to see how much improvement can be achieved by performing multiple sampling in the prediction step of the particle filter.

In Section 2, we briefly summarize Kalman filter, non-Gaussian filter and particle filters as filtering and smoothing algorithms for state-space models, and present some known results on the approximation accuracy of particle smoothing. In Section 3, we present a method for multiple sampling in the prediction step of the particle filter and show the improvement in accuracy and increase in computational complexity with increasing multiplicity $L$ for the same example problem as in Section 2. Section 4 discusses the impact of stratified sampling, and Section 5 summarizes the findings of this paper.

\section{A Brief Review of the Filtering and Smoothing Algorithms}

\subsection{The state-space model and the state estimation problems}

Assume that a time series \( y_n \) is expressed by a linear state-space model 
\begin{eqnarray}
x_{n} &=& F_nx_{n-1} \: + \: G_{n}v_{n} \label{ssm-1} \nonumber \\
y_n &=& H_nx_n \: + \: w_n, \label{ssm-1}
\end{eqnarray}
where \(x_n\) is an \( k \)-dimensional state vector,
\( v_{n} \) and \( w_n \) are \( \ell \)-dimensional and 1-dimensional white noise sequences having density functions \(q_{n}(v) \) and \( r_n(w) \), respectively.
The initial state vector \( x_0 \) is assumed to be distributed
according to the density \( p(x_0 ) . \)

The information from the observations up to time \( j \) is denoted by \(Y_j \),
namely, \( Y_j \equiv \{y_1,\ldots ,y_j\}\).
The problem of state estimation is to evaluate \( p(x_n|Y_j)\), 
the conditional density of \( x_n \)
given the observations \(  Y_j \) and the initial density \( p(x_0|Y_0) \equiv p(x_0) . \)
For \(n>j, n=j \) and \(n<j \), it is called the problem of prediction, filtering
and smoothing, respectively.\\

This linear state-space model can be generalized to a nonlinear non-Gaussian state-space model,
\begin{eqnarray}
 x_n & = & F_n(x_{n-1}, v_n) \nonumber \\ % \hspace{10mm}(System Model)\\
 y_n & = & H_n(x_n) + w_n, \label{ssm-2}% \hspace{11mm}(Observation Model)  \nonumber
\end{eqnarray}
where $F_n(x,v)$ and $H_n(x)$ are possibly nonlinear functions of the state and the noise inputs. Diverse problems in time series analysis can be treated by using this nonlinear state-space
model (Kitagawa and Gersch 1996, Doucet et al. 2001).
Note that this nonlinear non-Gaussian state-space model can be
further generalized to general state-space model which is defined by
using conditional distributions.

\subsection{The Kalman filter and the smoother}

It is well-known that if all of the noise densities \( q_n(v) \) and \( r_n(w) \) and the initial state density \( p(x_0) \) are Gaussian, then the conditional density of linear state-space model (1), 
\( p(x_n|Y_m) \), is also Gaussian and that the mean and the variance covariance matrix
can be obtained by the Kalman filter and the fixed interval smoothing algorithms (Anderson and Moore 1979).

To be specific, if we assume \( q_n(v) \sim N(0,Q_n) \), \(r_n(w) \sim N(0,R_n) \), \( p(x_0|Y_0) \sim N(x_{0|0},V_{0|0}) \) and \( p(x_n|Y_m) \sim N(x_{n|m},V_{n|m}) \), then the Kalman filter is given as follows:\\
{\bf One-step ahead prediction}:
\begin{eqnarray}
x_{n|n-1} &=& F_nx_{n-1|n-1} \nonumber \\
V_{n|n-1} &=& F_nV_{n-1|n-1}F_n^T + G_nQ_{n}G_n^T.
\end{eqnarray}
{\bf Filter}
\begin{eqnarray}
K_n     &=& V_{n|n-1}H_n^T(H_nV_{n|n-1}H_n^t + R_n)^{-1} \nonumber \\
x_{n|n} &=& x_{n|n-1} + K_n(y_n - H_nx_{n|n-1}) \\
V_{n|n} &=& (I - K_nH_n)V_{n|n-1}. \nonumber
\end{eqnarray}
 
Using these estimates, the smoothed density is obtained by the following,\ 

\noindent
{\bf Fixed interval smoothing algorithm}:
\begin{eqnarray}
A_n     &=& V_{n|n}F_n^TV_{n+1|n}^{-1} \nonumber \\
x_{n|N} &=& x_{n|n} + A_n(x_{n+1|N} - x_{n+1|n}) \\
V_{n|N} &=& V_{n|n} + A_n(V_{n+1|N} - V_{n+1|n})A_n^T. \nonumber
\end{eqnarray}

\subsection{The non-Gaussian filter and the smoother}

 It is well-known that for the nonlinear non-Gaussian state-space model (\ref{ssm-2}), 
the recursive formulas for obtaining the densities of the one step ahead predictor, the filter and the smoother are as follows:\\
{\bf One step ahead prediction:}
\begin{equation}
 p(x_{n} |Y_{n-1} )
  = \int_{- \infty}^\infty p(x_{n} | x_{n-1} )
p(x_{n-1} |Y_{n-1} )dx_{n-1}. 
\end{equation}
{\bf Filtering:}
\begin{equation}
 p(x_n |Y_n ) = \frac{p(y_n|x_n)p(x_n|Y_{n-1})}{\int p(y_n |x_n )p(x_n |Y_{n-1} )dx_n}.
\end{equation}
{\bf Smoothing:}
\begin{equation}
 p(x_n |Y_N ) 
  = p(x_n |Y_n ) \int_{- \infty}^\infty \frac{p(x_{n+1} |Y_N )
p(x_{n+1} |x_n )}{p(x_{n+1} |Y_n )}dx_{n+1} .
\end{equation}

 In Kitagawa (1987, 1988), an algorithm for implementing the non-Gaussian filter and smoother was developed by approximating each density function using a step-function or a continuous piecewise linear function  and by performing numerical computations.
This method was successfully applied to various problems such as estimation of trend
or volatility, spectrum smoothing, smoothing discrete process and tracking problem
(Kitagawa and Gersch 1996, Kitagawa 2020).

\subsection{Sequential Monte Carlo filter and smoother for non-Gaussian nonlinear state-space models}

The non-Gaussian filter and smoother based on numerical integration 
mentioned in the previous subsection has a limitation that
it can be applied to only lower dimensional, such as the  third or the fourth order, state-space model. 
Sequential Monte Carlo filter and smoother, hereinafter referred to as particle filter, were developed to mitigate
this problem.
In this method, each distribution appeared in recursive
filter and smoother is approximated by many ^^ ^^ particles"
that can be considered as realizations from that distribution
(Gordon et al. 1993, Kitagawa 1993, 1996).

In this paper, we use the following notations,
$ \{p_n^{(1)},\ldots , p_n^{(m)}\} \sim p(x_n|Y_{n-1})$, 
$ \{f_n^{(1)},\ldots , f_n^{(m)}\} \sim p(x_n|Y_{n})$, 
$ \{s_{n|N}^{(1)},\ldots , s_{n|N}^{(m)}\} \sim p(x_n|Y_{N})$.
In practice, we approximate the cumulative distributions by the empirical
distributions determined by the set of ^^ ^^ particles".

Then a recursive filtering algorithm is realized as follows:
% By this algorithm, a set of realizations expressing the one step ahead predictor $p(x_n|Y_{n-1})$ and the filter $p(x_n|Y_n)$ can be obtained recursively . Namely, $\{p_n^{(1)},\ldots ,p_n^{(m)}\}$ is obtained from $\{f_{n-1}^{(1)},\ldots ,f_{n-1}^{(m)}\}$ and subsequently $\{f_n^{(1)},\ldots ,f_n^{(m)}\}$ is from $\{p_n^{(1)},\ldots ,p_n^{(m)}\}$.
\begin{enumerate}
%\item {\rm Determine $m$, the number of realizations to be used for the approximation of each distribution.}
\item {\rm Generate a $k$-dimensional random number $f_0^{(j)} \sim p_0(x)$, for $j=1,\ldots ,m$.} 
\item {\rm Repeat the following steps for $n=1,\ldots ,N$}.
  \begin{enumerate}
  \item {\rm Generate an $\ell$-dimensional random number $v_n^{(j)} \sim q(v)$, for $j=1,\ldots ,m$}. 
  \item {\rm Generate a new particle by $p_n^{(j)} = F(f_{n-1}^{(j)},v_n^{(j)})$, for $j=1,\ldots ,m$.}
  \item {\rm Compute the importance weight $\alpha_n^{(j)} = r(y_n-H(p_n^{(j)}))$, of the particle $p_n^{(j)}$ 
for $j=1,\ldots ,m$.}
  \item {\rm Generate $f_n^{(j)} \sim (\sum_{i=1}^m\alpha_n^{(i)})^{-1} 
\sum_{i=1}^m \alpha_n^{(i)} I(x,p^{(i)}_n)$, for $j=1,\ldots ,m$ 
by the resampling of $p_n^{(1)},\ldots ,p_n^{(m)}$ with the sampling rate proportional to $\alpha_n^{(n)}$.}
  \end{enumerate}
\end{enumerate}

%\begin{figure}[tbp]
%\begin{center}
%\includegraphics[width=90mm,angle=0,clip=]{MCF-2.eps}
%\caption{Sequential Monte Carlo filter. One cycle of prediction,
%filter and resampling.}
%\label{figure:MCF}
%\end{center}
%\end{figure}

In Kitagawa (1993, 1996), it is shown that the particles approximating the 
smoothing distribution are obtained 
by a simple modification of the particle filter. 
Assume that $(s_{1|i}^{(j)},\ldots ,s_{n|i}^{(j)})^T$ denotes the $j$-th
realization of the conditional joint density $p(x_1,\ldots ,x_n|Y_i)$.
Then an algorithm for smoothing is obtained by replacing 
the Step 2 (d) of the algorithm for filtering; 

\vspace{2mm}

\hangindent=15mm 
(d-S) {\rm Generate $\{(s_{n-L|n}^{(j)},\cdots ,s_{n-1|n}^{(j)},
s_{n|n}^{(j)})^T , 
~j=1,\ldots ,m\}$ 
by the resampling of 
$\{(s_{n-L|n-1}^{(j)},$ $\cdots ,$ $s_{n-1|n-1}^{(j)},p_n^{(j)})^T,
~j=1,\ldots ,m\}$ with the sampling probability proportional to $\alpha_n^{(n)}$ }.

\vspace{2mm}

This is equivalent to applying the $L$-lag fixed lag smoother rather than the fixed interval smoother (Anderson and Moore 1979).
The increase of lag, $L$, will improve the accuracy of the 
$p(x_n|Y_{n+L})$ as an approximation to $p(x_n|Y_N)$, 
while it is very likely to decrease the accuracy of 
$\{s_{n|N}^{(1)}, \cdots, s_{n|N}^{(m)}\}$ as representatives of 
$p(x_n|Y_{n+L})$ (Kitagawa 1996). Since $p(x_n|Y_{n+L})$ usually converges quickly 
to $p(x_n|Y_N)$, it is recommended to 
take $L$ not so large. 

\subsubsection{Example: smoothing accuracy of trend model}
\begin{figure}[tbp]
\begin{center}
\includegraphics[width=80mm,angle=0,clip=]{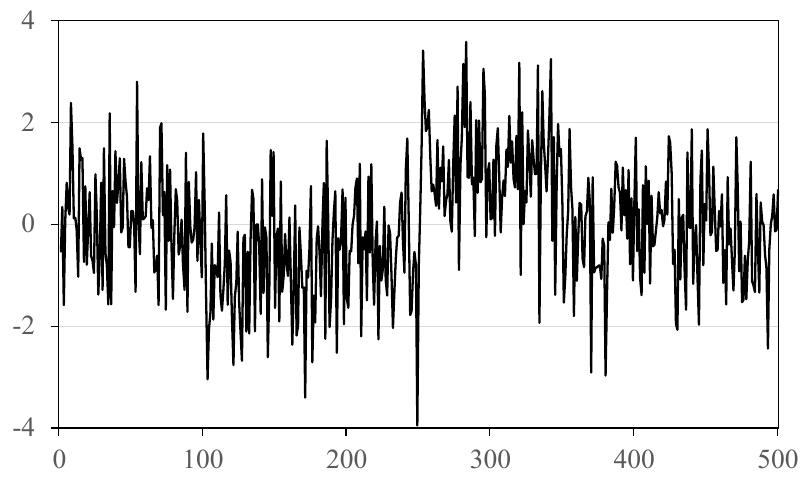}
\caption{Test data used for the Monte Carlo study, Kitagawa(1987, 1996, 2014)}
\label{figure:Test_data}
\end{center}
\end{figure}

For emperical study on the accuracy of the particle filter, we consider the following first-order trend models:
\begin{eqnarray}
  x_n &=& x_{n-1} + v_n \nonumber \\
  y_n &=& x_n + w_n,
\end{eqnarray}
where $y_n$ is the observed time series, $x_n$ is the trend component, $v_n$ is the system noise and $w_n$ is the observation noise.
It is assumed that the observation noise $w_n$ follows either the Gaussian distribution $N(0,\sigma^2)$ and
the system noise $v_n$ follows either the Gaussian distribution $N(0,\tau^2)$ or
the Cauchy distribution $C(0,\tau^2)$.
Figure \ref{figure:Test_data} shows the data used in the Monte Carlo experiments in this paper, which were also used by Kitagawa (1987, 1996, 2014).

As in Kitagawa (2014), the following quantities will be used as criteria for evaluating the estimated trends:
\begin{eqnarray}
  \mbox{Dist}(D,\hat{D}) = \sum_{n=1}^{500} \sum_{i=1}^{6400} \left\{ D(x_i,n) - \hat{D}(x_i,n)\right\}^2\Delta x, 
\end{eqnarray}
where $D(x_i,n)$  and $\hat{D}(x_i,n)$ are the ^^ ^^ true" and estimated probability distributions at time $n$ defined on the sampling points $x_i, i=1,\ldots ,6400$. $x_i, i=1,\ldots ,6400$ are defined by $x_i = -8+(i-1)\Delta x$ with $\Delta x = 16/6400$.
In actual evaluation, the $\hat{D}(x_i, n)$ is replaced with either the filter distribution $\hat{D}_f (x, n)$, or the smoother distribution $\hat{D}_s(x, n)$
obtained by the particle filter and $D(x_i , n)$ by the ^^ ^^ true" filter
distribution $D_f(x,n)$, or the fixed-interval smoother distribution $D_s(x,n)$, respectively.
For Gaussian distribution model, the true distribution is obtained by the Kalman filter. On the other hand,
for Cauchy distribution model, the ^^ ^^ true" distribution is obtained by numerical integration method (Kitagawa 1987).

\begin{figure}[tbp]
\begin{center}
\includegraphics[width=160mm,angle=0,clip=]{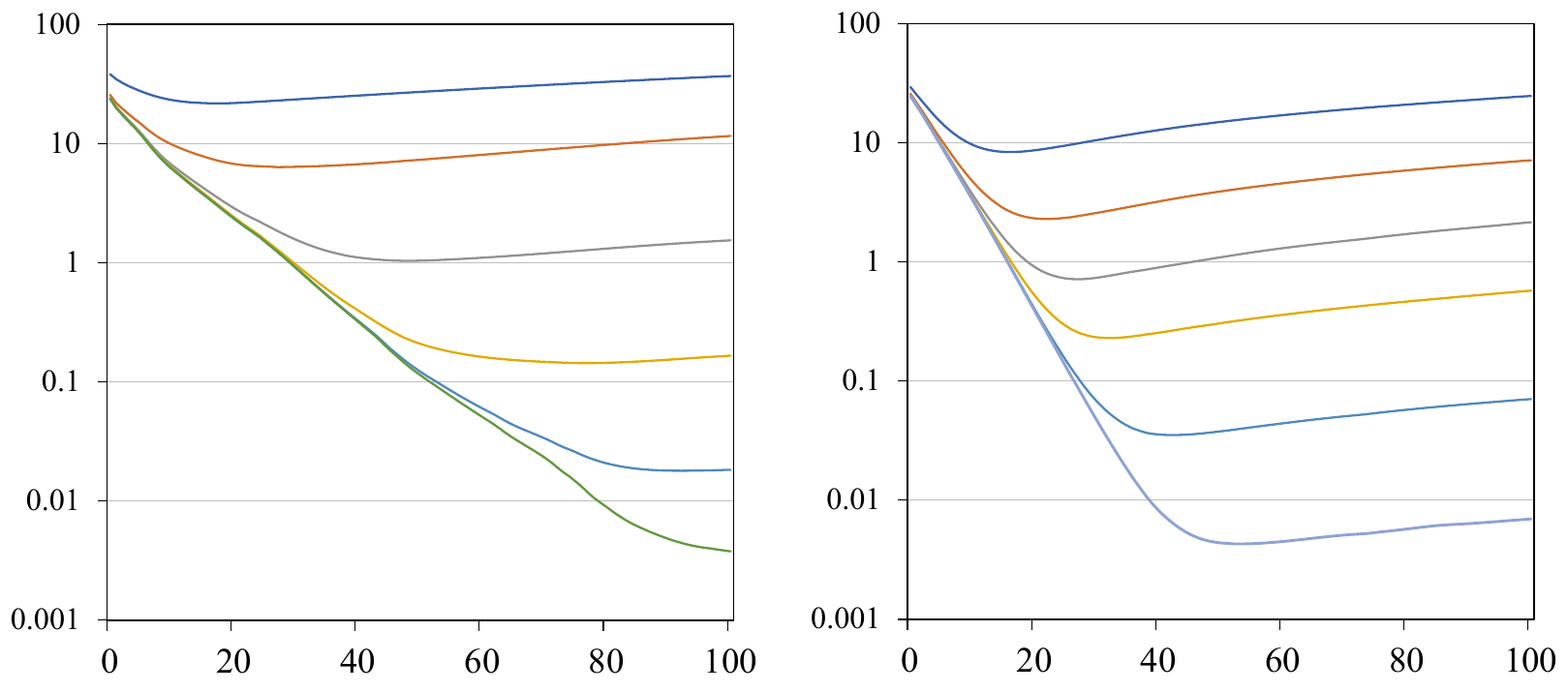}
\caption{Acuracy of smoother with LAG=$1,\ldots ,100$, number of particles $m=10^k$, $k=2,\ldots ,7$. Left plot: Cauchy model, right plot: Gaussian model. Figure X in Kitagawa (2014) has been redrawn. In the original figure, it is drawn as a double-logarithmic graph.}
\label{figure:Acuracy_vs_LAG}
\end{center}
\end{figure}

Figure \ref{figure:Acuracy_vs_LAG} reproduces the results of Kitagawa (2014), but note that the horizontal axis is not log(Lag), but Lag as is.
Table \ref{Tab_Accuracy_vs_LAG} shows the number of Lags for which the evaluation criterion Dist($D_s,\hat{D}_s$) is minimized for the number of particles $m=10^k$, $k=2, \ldots, 7$. Left plot shows the case of Cauchy noise models and the right plot show the case of Gaussian noise model.
In both models, the optimal Lag increases as the number of particles $m$ increases. In the case of the Gaussian noise model, the increase is less pronounced and about 50 is sufficient. In the case of the Cauchy noise model, on the other hand, the optimal lag increases with the number of particles $m$ faster than in the Gaussian model, reaching more than 100 for $m=10^7$.

\begin{table}[tbp]
\caption{Orders of lags that attain the minimum of the ^^ ^^ distance" of the estimated fixed-lag smoother and the ^^ ^^ true" fixed-interval smoother}
\label{Tab_Accuracy_vs_LAG}
\begin{center}
\begin{tabular}{c|cccccc}
$m$  & $10^2$ & $10^3$ & $10^4$ & $10^5$ & $10^6$ & $10^7$ \\
\hline
Gauss  & 16  &  22  &  27  &  32  &  43  &  53 \\
Cauchy & 17  &  28  &  48  &  80  &  93  & 108
\end{tabular}
\end{center}
\end{table}

\begin{figure}[tbp]
\begin{center}
\includegraphics[width=150mm,angle=0,clip=]{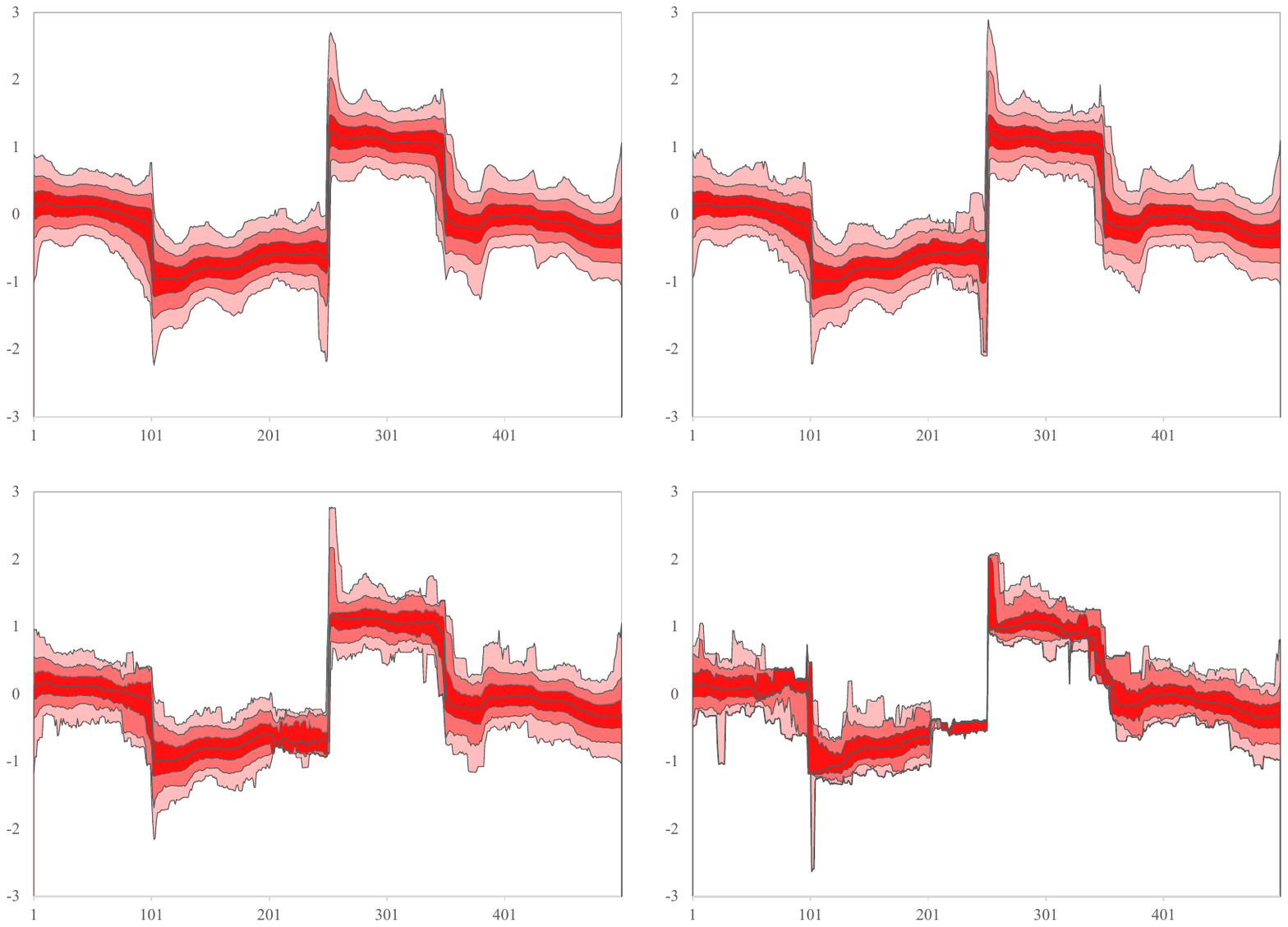}
\caption{Smoothed posterior distribution of the trend component by the standard particle smoother with $L=1$. From top left to bottom right, $m$=1,000,000, 100,000, 10,000 and 1,000.}
\label{figure:MCF with L=1}
\end{center}
\end{figure}

Figure \ref{figure:MCF with L=1} shows the posterior distribution of the trend component obtained by the standard
particle smoother for the Cauchy noise model with number of particles $m$=1,000,000, 100,000, 10,000 and 1,000.
The 0.13, 2.27, 15.87, 50.0, 84.13, 97.73 and 99.87\% points corresponding to the mean and $\pm$1, 2, and 3 standard deviations of the normal distribution are displayed. The ^^ ^^ true" smoothed distribution, estimated by non-Gaussian smoothing algorithm using numerical integration, is shown in Kitagawa (2014).
It can be seen that the distribution estimated by particle smoothing approaches the true smoothed distribution as the number of particles increases, and that fairly reasonable estimates of the 50\% point (median) are obtained even when $m = 1,000$.

%%%%%%%%%%%%%%%%%%%%%%%%%%%%%%%%%%%%%%%%%%%%%%%%%%%%%%%%%%%%%%%%%%%%
%\newpage
\section{Multi-Particle Prediction}
\subsection{Multi-sampling in Prediction Step}
Step 2.(a) of the particle filter algorithm usually generates one particle using one random number $v_n^{(j)}$, but in practice it is possible to generate two or more particles. Increasing the number of random numbers used to approximate the predictive distribution is expected to increase the accuracy of the approximation. If $L$ random numbers, $v_n^{(j,i)}, i=1,\ldots ,L$, are generated for each particle approximating the filter distribution of the previous time $n-1$, $f_{n-1}^{(j)}$, then the step 2. of the particle filter is modefied as follows:
\begin{enumerate}
  \item[(a)] {\rm Generate $L$ $\ell$-dimensional random numbers $v_n^{(j,i)} \sim q(v), i=1,\ldots ,L$, for $j=1,\ldots ,m$}. 
  \item[(b)] {\rm Generate $L$ new particles by $p_n^{(j,i)} = F(f_{n-1}^{(j)},v_n^{(j,i)}), i=1,\ldots,L$, for $j=1,\ldots ,m$.}
  \item[(c)] {\rm Compute the importance weight $\alpha_n^{(j,i)} = r(y_n-H(p_n^{(j,i)}))$, for $j=1,\ldots ,m$ and $i=1,\ldots ,L$.}
  \item[(d)] {\rm Generate $f_n^{(k)} \sim (\sum_{j=1}^m\sum_{i=1}^L\alpha_n^{(j,i)})^{-1} 
\sum_{j=1}^m\sum_{i=1}^L \alpha_n^{(j,i)} I(x,p^{(j,i)}_n)$, for $k=1,\ldots ,m$ 
by the resampling of $L\times m$ particles, $p_n^{(1,1)},\ldots ,p_1^{(1,L)},\ldots,p_n^{(m,1)},\ldots ,p_n^{(m,L)}$.}
\end{enumerate}

\subsection{Comparison with the Standard MCF ($L=1$)}

Table \ref{Tab:L_vs_filter_accuracy} shows the change in filter accuracy with the increase of random numbers $L$ used for prediction. The number of particles $m$ is considered for five types of $10^k$, $k=2,\ldots,6$, and $L$ is from 1 to 10. The table shows the average of $D(f;\hat{f})$'s obtained by NSIM times of filtering with different random numbers. NSIM is set to 1,000 for $m=10^2$ and $10^3$, 400 for $m=10^4$, 100 for $m=10^5$ and 25 for $m=10^6$. The four columns on the left represent filter accuracy of the Gaussian noise model and the five columns on the right represent the results for the Cauchy noise model.

\begin{table}[tbp]
\caption{Change of filter accuracy with the number of random numbers $L$ ($L=1,\ldots ,10$) for various numbers of particles, $m=$100, 1,000, 10,000 and 100,000. The table shows the average of 100 calculations using different random numbers. Left: Gaussian model, right: Cauchy model.}
\label{Tab:L_vs_filter_accuracy}
\begin{center}
\begin{tabular}{c|cccc|ccccc}
     &\multicolumn{4}{c|}{Gauss}&\multicolumn{5}{c}{Cauchy}\\
 $L$ & 100 &1,000& $10^4$&$10^5$  & 100& 1,000& $10^4$ &$10^5$&$10^6$\\
\hline
    1&3.822&0.592&0.125 & 0.033  & 22.771 & 4.863 & 0.387 & 0.033 & 0.0038\\
    2&3.274&0.512&0.100 & 0.020  & 17.735 & 2.803 & 0.235 & 0.024 & 0.0025\\
    3&3.182&0.480&0.093 & 0.021  & 15.443 & 2.156 & 0.196 & 0.020 & 0.0022\\
    4&3.076&0.471&0.096 & 0.020  & 13.863 & 1.826 & 0.175 & 0.018 & 0.0021\\
    5&2.983&0.455&0.088 & 0.017  & 12.791 & 1.666 & 0.171 & 0.017 & 0.0019\\
    6&3.009&0.450&0.081 & 0.019  & 11.986 & 1.518 & 0.159 & 0.017 & 0.0017\\
    7&2.964&0.445&0.091 & 0.021  & 11.527 & 1.458 & 0.152 & 0.017 & 0.0018\\
    8&2.981&0.434&0.092 & 0.028  & 10.860 & 1.427 & 0.147 & 0.016 & 0.0017\\
    9&2.962&0.445&0.082 & 0.015  & 10.485 & 1.374 & 0.151 & 0.015 & 0.0017\\
   10&2.938&0.432&0.085 & 0.016  & 10.285 & 1.349 & 0.151 & 0.015 & 0.0015\\
\hline
\end{tabular}
\end{center}
\end{table}

\begin{figure}[tbp]
\begin{center}
\includegraphics[width=120mm,angle=0,clip=]{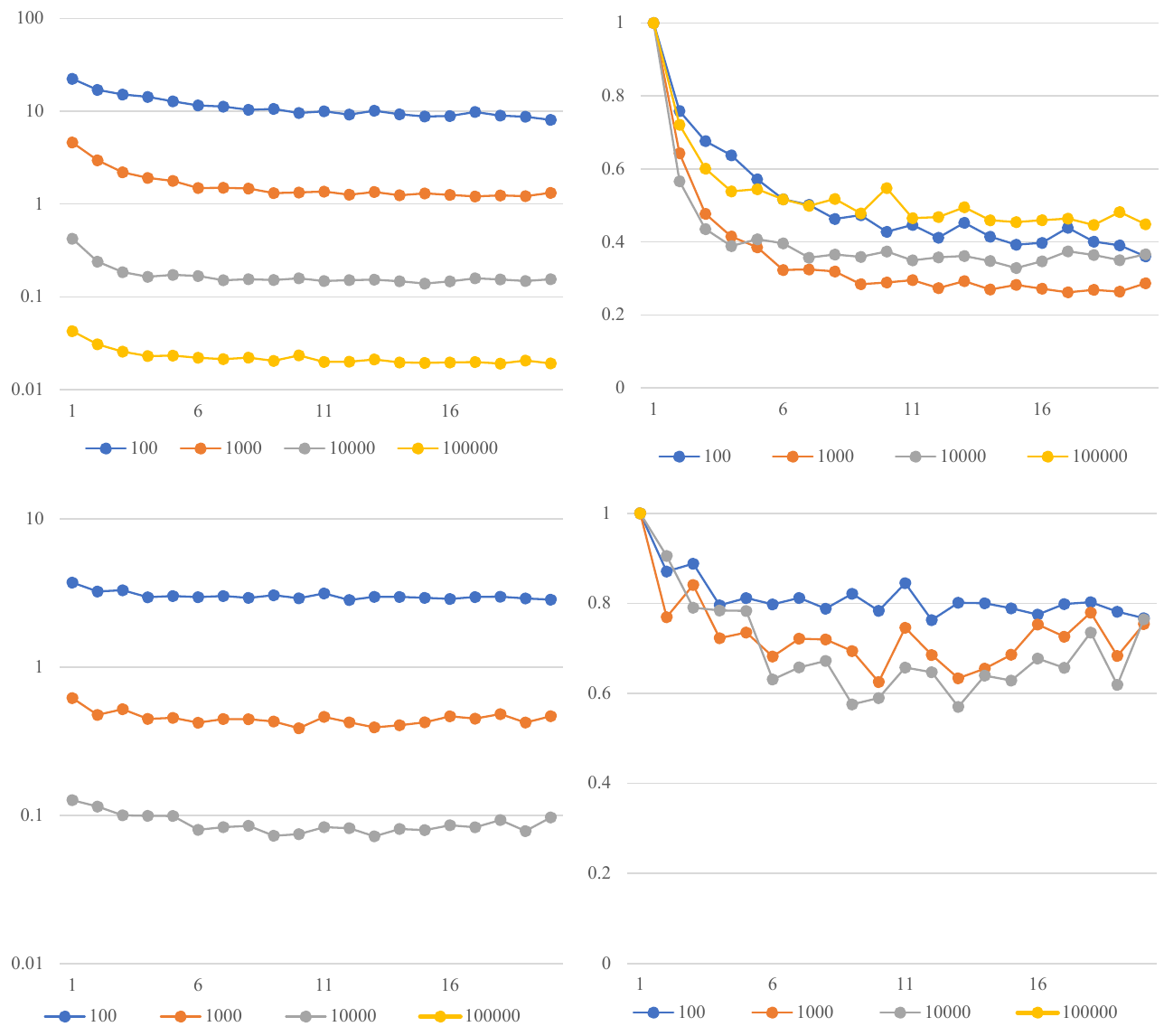}
\caption{Changes of filter error variances for $m=100$, 1,000 and 10,000. 
Upper plots: Cauchy model, lower plots: Gaussian model.
Left plots: variances in log-scale, right plots: decrease ratio relative to the case $L=1$. }
\label{figure:L_vs_filter_variance}
\end{center}
% \end{figure}
% \begin{figure}[tbp]
\begin{center}
\includegraphics[width=150mm,angle=0,clip=]{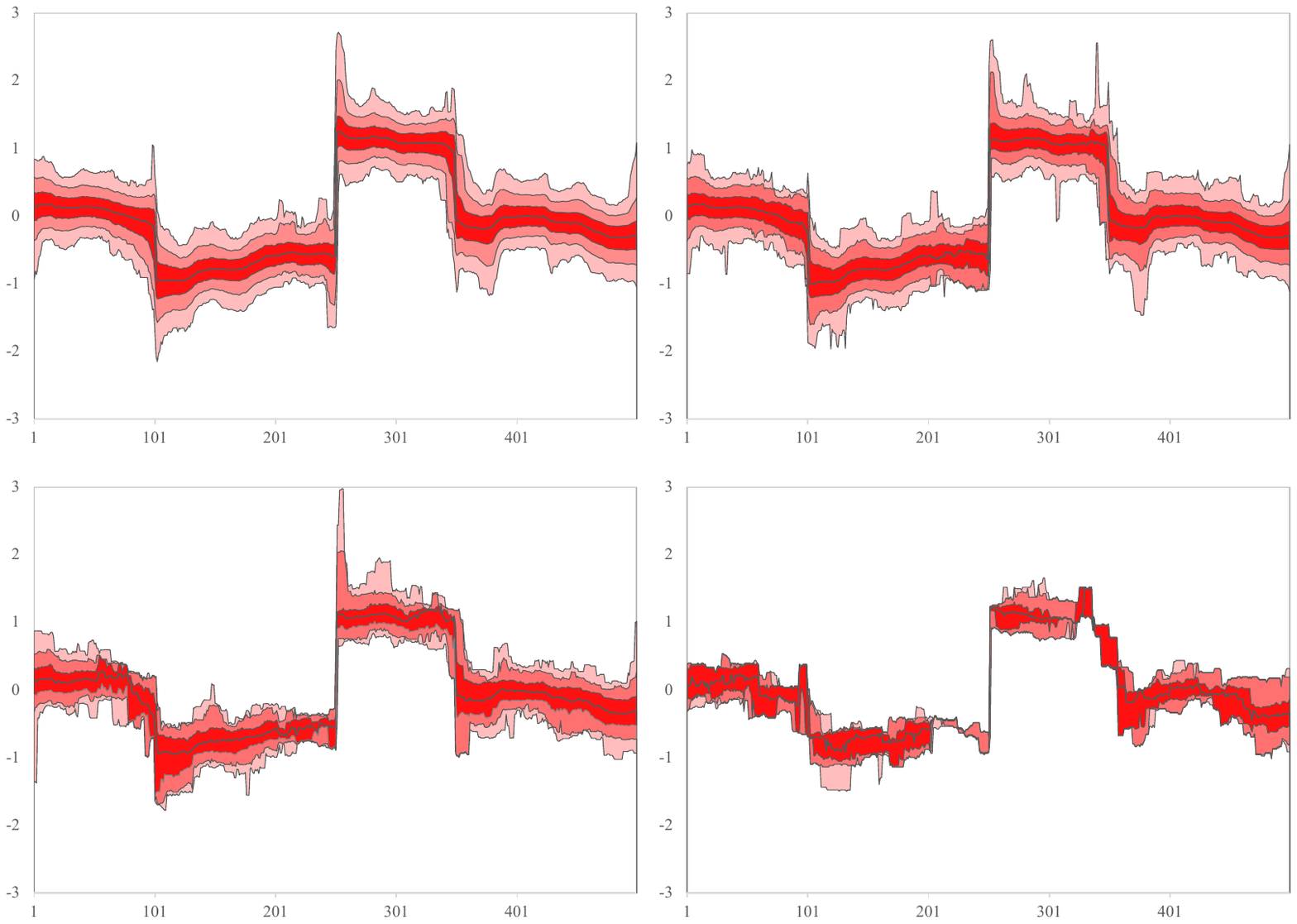}
\caption{MCF with $L=10$. From top left to bottom right $m=100,000$, 10,000, 1,000 and 100.}
\label{figure:MCF with L=10}
\end{center}
\end{figure}

Figure \ref{figure:L_vs_filter_variance} illustrates the results of Table \ref{Tab:L_vs_filter_accuracy}, the upper plots for the Cauchy noise model and the lower plots for the Gaussian noise model. The left plots show the accuracy on a logarithmic scale, and the right plots show how much the approximation error of the distribution is reduced for each number of particles, compared to the simple particle filter with $L=1$.

In both cases, the reduction in error is not significant above $L=5$, indicating that $L$ at most 5 is sufficient. It should be noted, however, that the error for the Cauchy noise model is reduced to 30\% to 50\%, while for the Gaussian noise model it is only 60\% to 80\%, indicating that the effect of using a large $L$ is not significant for the Gaussian noise model.

Figure \ref{figure:MCF with L=10} shows the smoothed posterior distribution of the Cauchy model obtained with $m=10^k$, $k=5$, 4, 3 and 2, and $L=10$. Note that the number of particles are $1/10$ of that in Figure \ref{figure:MCF with L=1} where the results for $m=10^k$, $k=6$, 5, 4 and 3 are shown. It can be seen that the smoothed posterior distribution obtained with $m$ particles and $L=10$ have an accuracy intermediate between that obtained with $m$ particles and $L=1$ and that obtained with $10m$ particles and $L=1$. This is confirmed by the values in Table \ref{Tab:L_vs_filter_accuracy}.

%\begin{figure}[h]
%\begin{center}
%\includegraphics[width=160mm,angle=0,clip=]{MCF_with_L=100.pdf}
%\caption{Standard MCF with $L=100$.}
%\label{figure:MCF with L=100}
%\end{center}
%\end{figure}

%\begin{figure}[tbp]
%\begin{center}
%\includegraphics[width=80mm,angle=0,clip=]{MCF_with_various_L.pdf}
%\caption{MCF with more than one system noises. Horizontal axis: $L$.}
%\label{figure:MCF with L>1}
%\end{center}
%\end{figure}

\subsection{CPU-times for Multi-particle Prediction}
Table \ref{Tab:L_vs_CPU-time} shows the increase in CPU time in second when the number of random numbers generated for each particle of filter is increased to $L=1,\ldots ,10$. The number of particles is $m=10^k$, $k=2,\ldots ,6$.
The left half of the table is for the filter case (Lag=0) and the right half is for the smoothing case with Lag=100.
The CPU time was measured NSIM times on a Windows PC (Intel Core i7-8700, 3.20GHz, 32GB RAM), and the average is shown.

For the case of Lag=0, filtering with $L$=10 requires about 8 times more CPU time than with $L$=1, while for the case of Lag=100, it requires only about 3 times more CPU time. This is because the time required to exchange stored particles in the filter (and smoothing) step does not increase the computational complexity even with $L$=10.

\begin{table}[h]
\caption{Change of CPU-time with the number of random numbers $L$ ($L=1,\ldots ,20$) for various numbers of particles, $m=$100, 1,000, 10,000 and 100,000 for Cauchy model. Left: Filter (Lag=0), right: Smoother with Lag=100.}
\label{Tab:L_vs_CPU-time}
\begin{center}
\begin{tabular}{c|ccccc|ccccc}
     &\multicolumn{5}{c|}{Lag=0}&\multicolumn{5}{c}{Lag=100}\\
 $L$ & 100 &1,000&10,000&$10^5$ & $10^6$ & 100& 1,000&10,000&$10^5$&$10^6$\\
\hline
1 &	0.004&	0.039&	0.385&	3.095&	38.786&		0.018&	0.168&	1.286&	18.389&	181.875\\
2 &	0.006&	0.061&	0.622&	5.320&	74.016&		0.022&	0.193&	1.835&	23.506&	221.012\\
3 &	0.008&	0.091&	0.846&	7.881&	95.113&		0.025&	0.219&	2.441&	24.515&	262.762\\
4 &	0.013&	0.119&	1.161&	11.005&	129.009&	0.030&	0.243&	2.949&	41.217&	297.918\\
5 &	0.014&	0.157&	1.139&	14.802&	191.639&	0.030&	0.292&	3.292&	42.184&	363.965\\
6 &	0.017&	0.181&	1.357&	16.688&	197.459&	0.036&	0.331&	3.149&	45.131&	417.066\\
7 &	0.018&	0.213&	1.976&	20.548&	218.151&	0.040&	0.348&	3.533&	41.389&	447.738\\
8 &	0.027&	0.238&	2.116&	23.077&	241.167&	0.044&	0.444&	4.346&	40.310&	459.441\\
9 &	0.029&	0.271&	2.137&	26.380&	257.848&	0.044&	0.483&	4.353&	56.978& 534.910\\
10&	0.027&	0.318&	2.553&	27.529&	268.546&	0.047&	0.519&	4.686&	51.445& 535.230\\
\hline
\end{tabular}
\end{center}
%\end{table}
%\begin{table}[tbp]
\caption{Change of CPU-time for various numbers of particles, $m=m_0\times l_0$ for $m_0=$100, 1,000, 10,000, 100,000 and $l_0=1,\ldots ,10$ for Cauchy model. Left: Filter (Lag=0), right: Smoother with Lag=100.}
\label{Tab:L_vs_CPU-time2}
\begin{center}
\begin{tabular}{c|ccccc|ccccc}
     &\multicolumn{5}{c|}{Lag=0}&\multicolumn{5}{c}{Lag=100}\\
 $l_0$ & 100 &1,000&10,000&$10^5$ & $10^6$& 100& 1,000&10,000&$10^5$&$10^6$\\
\hline
1&	0.003&	0.034&	0.372& 	4.256&	37.157&	0.015 &	0.172 &	1.882 &	20.893&	196.918 \\
2&	0.008&	0.069&	0.721& 	7.418&	82.883&	0.032 &	0.342 &	3.889 &	47.013&	397.143 \\
3&	0.011&	0.101&	1.063& 	10.665&	122.339&0.048 &	0.554 &	5.729 &	67.748&	600.647 \\
4&	0.014&	0.141&	1.397& 	14.787&	174.113&0.062 &	0.760 &	8.552 &	113.456&846.428 \\
5&	0.018&	0.176&	1.928& 	17.757&	204.140&0.083 &	0.950 &	9.393 &	140.603&1080.004 \\
6&	0.022&	0.215&	2.447& 	23.755&	249.149&0.098 &	1.069 &	12.842&	160.852 &1307.047 \\
7&	0.026&	0.257&	3.303& 	25.455&	288.023&0.108 &	1.239 &	14.533&	182.767 &1376.278 \\
8&	0.029&	0.328&	3.940& 	27.676&	313.691&0.125 &	1.484 &	14.796&	168.385 &1694.059 \\
9&	0.033&	0.389&	4.263& 	32.157&	360.874&0.143 &	1.893 &	16.988&	265.403 &1900.401 \\
10&	0.035&	0.378&	4.654& 	42.685&	469.183&0.184 &	1.909 &	21.040&	196.918 &2167.504 \\
\hline
\end{tabular}
\end{center}
\end{table}

Table \ref{Tab:L_vs_CPU-time2} shows the results when the number of particles $m$ used for filtering is increased as $m = m_0\times l_0$. In this case, the CPU time for both filtering (Lag=0) and smoothing (Lag=100) increased more than 10 times when $l_0$=10 compared to the case with $l_0$=1.

\begin{figure}[tbp]
\begin{center}
\includegraphics[width=150mm,angle=0,clip=]{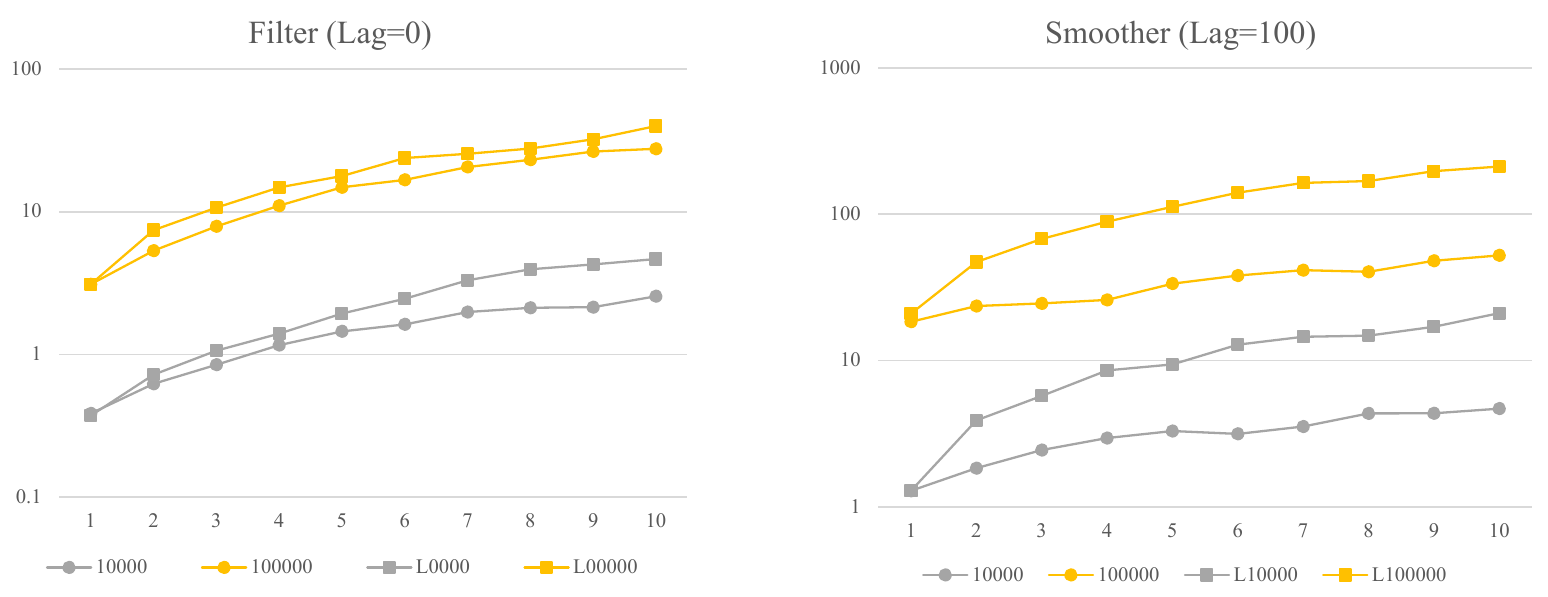}
\caption{Comparison of CPU-time for $m$, $L=1,\ldots ,10$ (circle) and $m, 2\times m, \ldots, 10\times m$ (square), $m=10,000$ (gray) and 100,000 (orage). Left plot: Filter (Lag=0), right plot: Smoother with Lag=100. Cauchy noise model case. Horizontal axis: $L$.}
\label{figure:CPU_time}
\end{center}
\end{figure}

Figure \ref{figure:CPU_time} illustrates the results of Table \ref{Tab:L_vs_CPU-time2} only for the cases $m = 10^4$ and $10^5$. {\footnotesize $\blacksquare$} shows the case where the number of particles is simply increased, and {\large $\bullet$} shows the case of multi-particle prediction where the number of particles is increased only in the prediction step without increasing the number of particles in the filter step. The left plot is for LAG=0, and the right plot is for Lag=100.
In the case of Lag=0, the decrease in CPU time is not so significant by multi-particle prediction. This is confirmed by Figure 7, which shows Computational Efficiency defined as (CPU-time of multi-particle prediction)/(CPU-time of single-particle prediction). On the other hand, in the case of smoothing with Lag=100, multi-particle prediction reduces the computation time to about 30\%$\sim$20\% for $L>3$.

\begin{figure}[h]
\begin{center}
\includegraphics[width=160mm,angle=0,clip=]{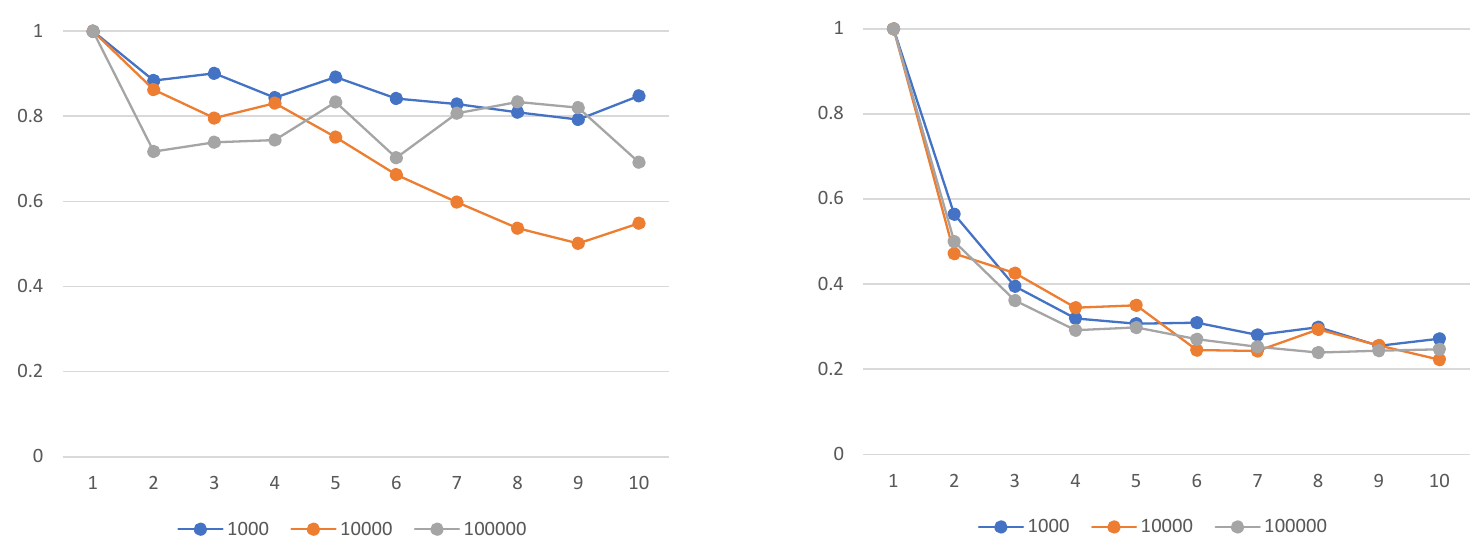}
\caption{Efficiency of multi-particle prediction. Left plot: filter, Right plot: smoother with LAG=100}
\label{figure:Efficiency_multi-particle}
\end{center}
%\end{figure}
%\begin{figure}[tbp]
\begin{center}
%\includegraphics[width=75mm,angle=0,clip=]{L_vs_Lm.pdf}
%\hspace{2mm}
%\includegraphics[width=79mm,angle=0,clip=]{L_vs_Lm2.pdf}
\includegraphics[width=160mm,angle=0,clip=]{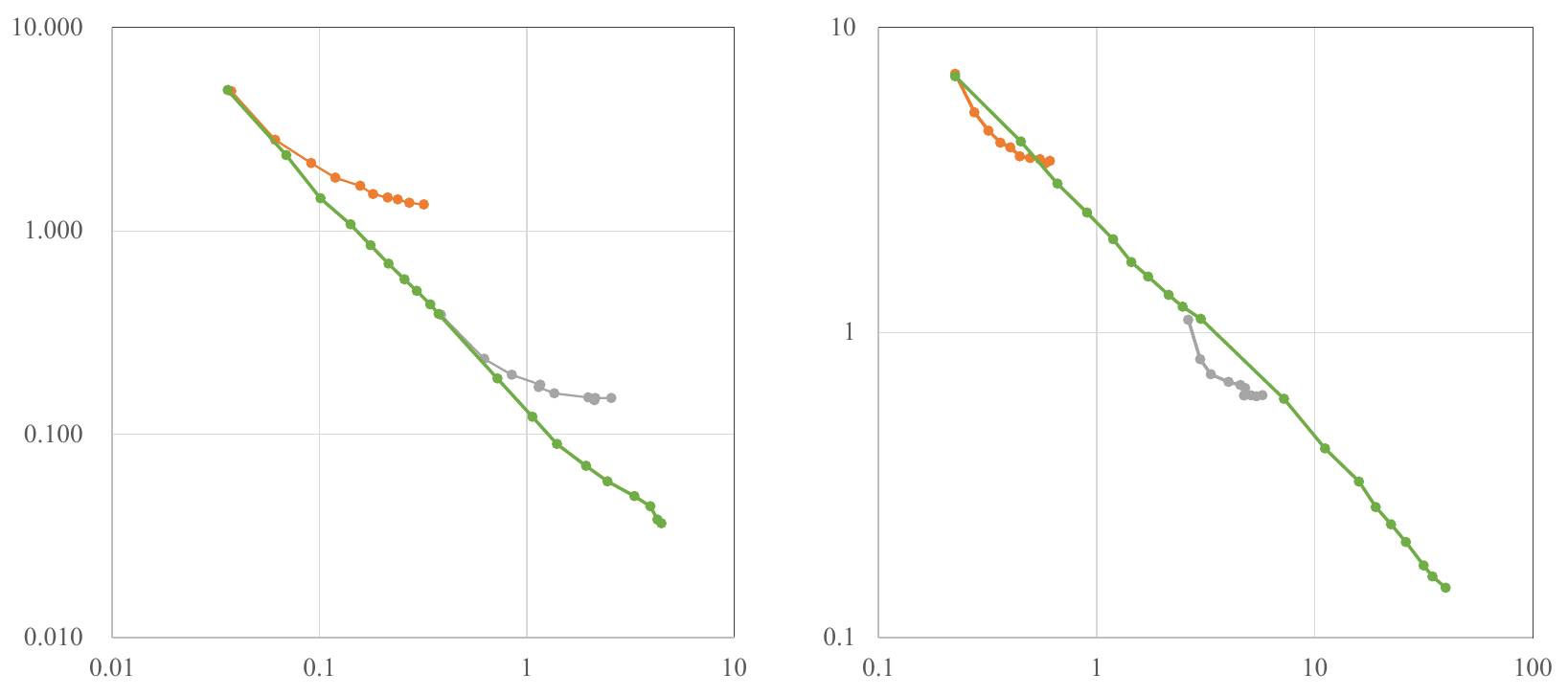}
\caption{Scatter plot of Accuracy vs. CPU-time.}
\label{figure:L_vs_Lm}
\end{center}
\end{figure}

Figure \ref{figure:L_vs_Lm} shows scatter plots with CPU time on the horizontal axis and accuracy of the particle filter or smoother on the vertical axis. The plots on the left are for the filter (Lag=0), while the plots on the right are for the smoother (Lag=100). The orage and gray lines show the cases, $L=1,\ldots ,10$, $m=10^4$ and $10^5$, respectively. On the other hand, green line shows case that $m=m_0\times l_0$, where $m_0=10^4$ and $10^5$, $l_0=1,\ldots ,10$, and $L=1$.
The area below the green line indicates that the multi-particle method is more efficient than the simple method in terms of CPU time. From the left plot, there is no advantage to use $L$ larger than 1 in the case of filter (Lag=0). On the other hand, in the case of smoothing (LAG=100), it can be seen that multi-particle smoothing with $L$ larger than 2, such as $2,\ldots 6$, is more efficient than the ordinary smooothing.
It should be noted that this increase in accuracy can be achieved without increasing the memory required for the smoothing algorithm.

\newpage
%%%%%%%%%%%%%%%%%%%%%%%%%%%%%%%%%%%%%%%%%%%%%%%%%%%%%%%%%%%%%%%%%%%%
\section{Balanced and Stratified System Noise}

In the multiple sampling described in the previous section, $L$ particles, $v_n^{(j)}, j=1,\ldots ,L$, were randomly generated according to the Gaussian or Cauchy distribution, but it is expected that a better approximation can be obtained by balancing, such as setting the mean to 0, or by performing stratified sampling, especially for small values of $L$.

\subsection{Balanced system noises}

By setting the mean (i.e., the first-order moment) of the $L$ random numbers to 0, the approximation accuracy of the MCF may be increased. For example, such system noise inputs can be generating by, for $L=2$, $v_n^{(j,1)}\sim q(v)$, $v_n^{(j,2)} =-v_n^{(j,1)}$, and for $L=3$, $r \sim q(v), s\sim q(v)$, $v_n^{(j,1)} = |r|$,
$v_n^{(j,2)}=-|s|$, $v_n^{(j,3)} = -(v_n^{(j,1)}+v_n^{(j,2)})$.
However, with this ad hoc method, it is difficult to generate the particles for $L=4$ or more.

%\begin{table}[h]
%\caption{Blanced sampling}\label{Tab_balancced sampling}
%\begin{center}
%\begin{tabular}{cc|ccccc}
%Model & $L$ & 100  & 1,000  & $10^4$ & $10^5$ & $10^6$ \\
%\hline
%Cauchy & 2  & 18.112 & 3.122 & 0.233 & 0.029 & 0.0077 \\
%       & 3  & 13.820 & 2.243 & 0.417 & 0.263 & 0.2437 \\
%\hline
%Gauss  & 2  &  3.260 & 0.479 & 0.099 & 0.014 & 0.0019 \\
%       & 3  &  3.411 & 0.661 & 0.214 & 0.098 & 0.0719 \\
%\hline
%\end{tabular}
%\end{center}
%\end{table} 

%Table \ref{Tab_balancced sampling} shows the accuracy of filter evaluated by $Dist(D_f,\hat{D}_f)$ for
%$L=2$ and 3. 

\subsection{Stratified system noises}

A method that can generate random numbers more systematically than balanced sampling is stratified sampling.
The stratified sampling in random number generation is considered here.
In the generation of $L$ system noises,
\begin{enumerate}
\item (a-S) Generate $i$-th system noise by $v_n^{(j,i)}=\Phi^{-1}(r_i)$, 
where $\Phi^{-1}(r)$ is the inverse function of the distribution function of the
system noise $v_n$ and $r_i$ is the uniform random number on $(u_0,u_1)$ obtained by 
$r_i \sim U(u_0, u_1)$ with $u_0 = (i-1)/L$ and $u_1=i/L$.
\end{enumerate}

\begin{figure}[tbp]
\begin{center}
\includegraphics[width=90mm,angle=0,clip=]{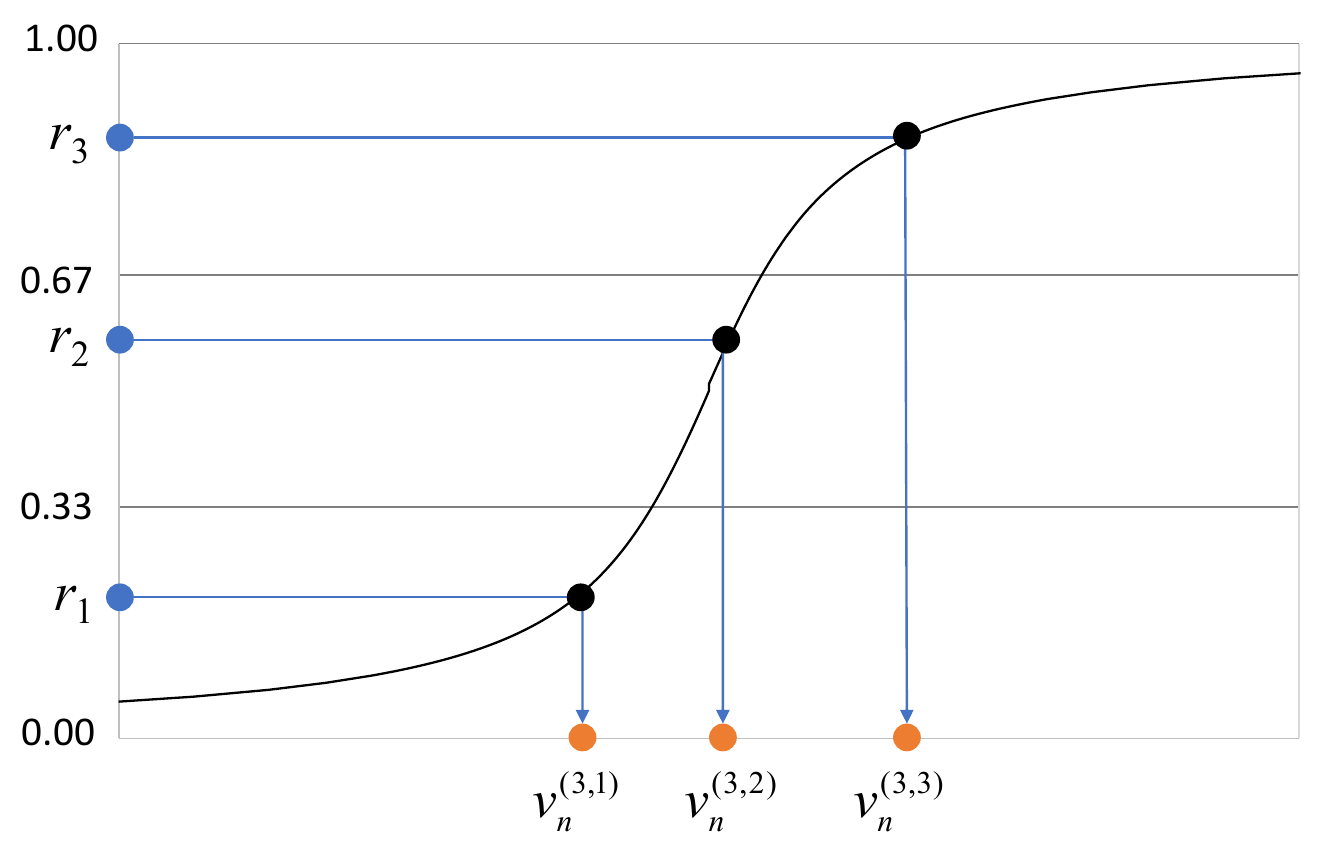}
\caption{Example of generating stratified sampling for $L$=3.}
\label{figure:Stratifed sampling}
\end{center}
\end{figure}

Figure \ref{figure:Stratifed sampling} shows an example of stratified sampling for $L$=3. The vertical axis shows the uniform random number $r$. The [0,1] interval is divided into three layers [0,1/3), [1/3,2/3), [2/3,1], and the random numbers $r_1$, $r_2$, $r_3$ are generated from each layer. Then, stratified random numbers can be generated by $v_n^{(3,i)}=\Phi^{-1}(r_i)$ using the inverse function of the noise distribution. Note that, if the noise distribution is Cauchy distribution, etc., the inverse function can be easily obtained as $\Phi(r) = \tan (\pi r)$.

\begin{figure}[tbp]
\begin{center}
\includegraphics[width=60mm,angle=0,clip=]{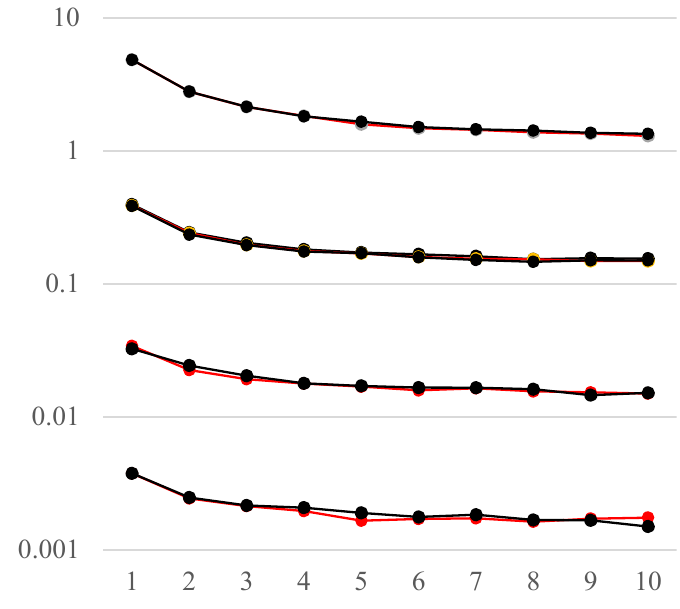}
\caption{Comparison of accuracy of the filter with stratified system noise and simple MCF.From top to bottom: $m=10^3, 10^4, 10^5$ and $10^6$. Red curve: stratified sampling, black curve: random sampling.}
\label{figure:Comarison of stratifed system noises and simple MCF}
\end{center}
\end{figure}

Table \ref{Table:Stratified_sampling} shows the comparison between stratified and random sampling for $m=10^k$, ($k=2,\ldots ,6$), $L=1,\ldots ,10$ for the test data shown in Figure 1. These numbers represent the average of accuracy evaluated by DIST$(D_f,\hat{D}_f)$ when NSIM times of the particle filtering wer performed using different random numbers. Here, NSIM=1,000 for $m=10^2$, $10^3$ and $10^4$ and NREP=100 for $m=10^5$ and NREP=25 for $10^6$.
Figure \ref{figure:Comarison of stratifed system noises and simple MCF} illustrates the same results for $m=10^3, 10^4, 10^5$ and $10^6$.
The results for the stratified sampling shown in red curves and the random sampling shown in black curves.
The two curves almost overlap, unexpectedly, indicating that the accuracy can hardly be cimproved by using stratified sampling.

\begin{table}[tbp]
\caption{Effect of stratified sampling}
\label{Table:Stratified_sampling}
\begin{center}
\begin{tabular}{c|ccccc|ccccc}
     &\multicolumn{5}{c|}{Stratified sampling} &\multicolumn{5}{c}{Random sampling} \\
     &\multicolumn{5}{c|}{Number of particles, $m$} &\multicolumn{5}{c}{Number of particles, $m$} \\
$L$   & 100 & 1,000 & $10^4$ & $10^5$ & $10^6$  & 100 & 1,000 & $10^4$ & $10^5$ & $10^6$ \\
\hline
1 &22.771	&4.863 &	0.390   &0.034  & 0.0038 & 22.771 & 4.863 & 0.387 & 0.033 & 0.0038\\
2 &18.268	&2.803 &	0.242	&0.023  & 0.0024 & 17.735 & 2.803 & 0.235 & 0.024 & 0.0025\\
3 &15.308	&2.152 &	0.197	&0.019  & 0.0021 & 15.443 & 2.156 & 0.196 & 0.020 & 0.0022\\
4 &13.913	&1.838 &	0.177	&0.018  & 0.0020 & 13.863 & 1.826 & 0.175 & 0.018 & 0.0021\\
5 &12.842	&1.593 &	0.170	&0.017  & 0.0017 & 12.791 & 1.666 & 0.171 & 0.017 & 0.0019\\
6 &12.067	&1.487 &	0.160	&0.016  & 0.0019 & 11.986 & 1.518 & 0.159 & 0.017 & 0.0017\\
7 &11.094	&1.448 &	0.155	&0.016  & 0.0017 & 11.527 & 1.458 & 0.152 & 0.017 & 0.0018\\
8 &10.893	&1.386 &	0.154	&0.016  & 0.0015 & 10.860 & 1.427 & 0.147 & 0.016 & 0.0017\\
9 &10.552	&1.356 &	0.149	&0.015  & 0.0015 & 10.485 & 1.374 & 0.151 & 0.015 & 0.0017\\
10&10.330	&1.301 &	0.149	&0.015  & 0.0017 & 10.285 & 1.349 & 0.151 & 0.015 & 0.0015\\
\hline
\end{tabular}
\end{center}
\end{table}

\newpage
\section{Conclusion}

An empirical study on the effect of using multi-particle prediction in particle filtering and smoothing revealed the following:
\begin{enumerate}
\item Smoothing with Lag=100 for $L>3$ by multi-particle prediction reduces computation time by about 70\% compared to $L$ times the number of particles, but only by about 20\% for the filter.
\item As seen in Figure 8, there is little merit in using the multi-particle prediction in the case of the filter (Lag=0) in terms of computational efficiency. However, in the case of smoothing, there are two advantages of multi-particle prediction:
\begin{enumerate}
  \item The computational efficiency is improved for $L<7$ or so.
  \item In the case of smoothing, a large memory, $m\times k\times$LAG$\times n$, is required, where $k$ is the state-dimension, $n$ is the number of data. However, in the case of multi-praticle prediction, the accuracy of the smoothed distribution can be improved without increasing the memory required for smoothing.
\end{enumerate}
\item At least as far as the data used in the empirical study are concerned, stratified sampling had no effect on improving accuracy at all.
\end{enumerate}

\end{document}